\documentclass{mem}
\usepackage{natbib}
\usepackage{txfonts}
\usepackage{balance}
\usepackage{graphicx}
\usepackage{epsfig}

\usepackage{epstopdf}
\idline{75}{282}
\begin{document}

\title{
Is dynamic heating of stellar disk inevitable?}

   \subtitle{}

\author{
A. \,Zasov\inst{1}, 
A. \, Saburova\inst{1}
   \and     I. \, Katkov\inst{1}
   }

  \offprints{A. Zasov}

\institute{
Sternberg Astronomical Institute of Moscow State University, \\ Universitetskii pr. 13, 119992, Moscow, Russia 
\email{zasov@sai.msu.ru}
}

\authorrunning{Zasov}

\titlerunning{ Is dynamic heating of stellar disk inevitable?}

\abstract{
Major mergers or/and the repeated minor mergers lead to dynamical heating of disks of galaxies. We analyze the available data on the velocity dispersion of stellar disks of S-S0 galaxies, including the new observational data obtained at 6m telescope of SAO RAS. As a measure of dynamical (over)heating, we use the ratio of the observed velocity dispersion to the minimal dispersion which provides the local  stability of the stellar disks with respect to gravitational perturbations. We came to conclusion that stellar disks in a significant part of galaxies (including LSB and some S0 galaxies) are close to the marginal stability condition (or are slightly overheated) -- at least at radial distances $r\sim$ 2-3 radial scalelenghts. It enables to constrain the role of merging in the heating of stellar disks: in many cases it seems to be non-efficient.  Marginal stability condition may also be successfully used to estimate the mass of a disk and the midplane volume gas (stars) densities on the basis of kinematic measurements.
}
\maketitle{}

\section{Introduction}

 The generally accepted point of view claims that a significant part of the progenitors of presently observed disky galaxies have experienced a major merging which could easily destroy stellar disks or at least heat them up dynamically. Minor mergers may also play a significant role. The question is: how really ''hot'' are the stellar disks? 

We collect the available information and get new observational data on the velocity dispersion of old stellar disks at $r/h >2$, where $h$ is the disk radial scalelength, and compare it with the minimal values of velocity dispersion which guarantee the local gravitational stability of disks. 

Radial stellar velocity dispersions of disks at different radial distances $r$ are compared with the critical value $(c_r)_{cr}$ expected for marginally stable 3D disks: $(c_r)_{cr}=Q_T\cdot c_T$, where $c_T=3.36G \sigma / \kappa$.
Here $\kappa$ is the epicyclic frequency, $\sigma$ is the disk surface density, $c_r$ is the radial velocity dispersion. Radial profile of the Toomre parameter $Q_T(r/h)$ was taken from the numerical simulations of \citet{Khoperskov2003}. It changes along radius between $\sim 1.2$ and 3, growing outside.

\begin{figure}[]
\resizebox{7.5cm}{!}{\includegraphics{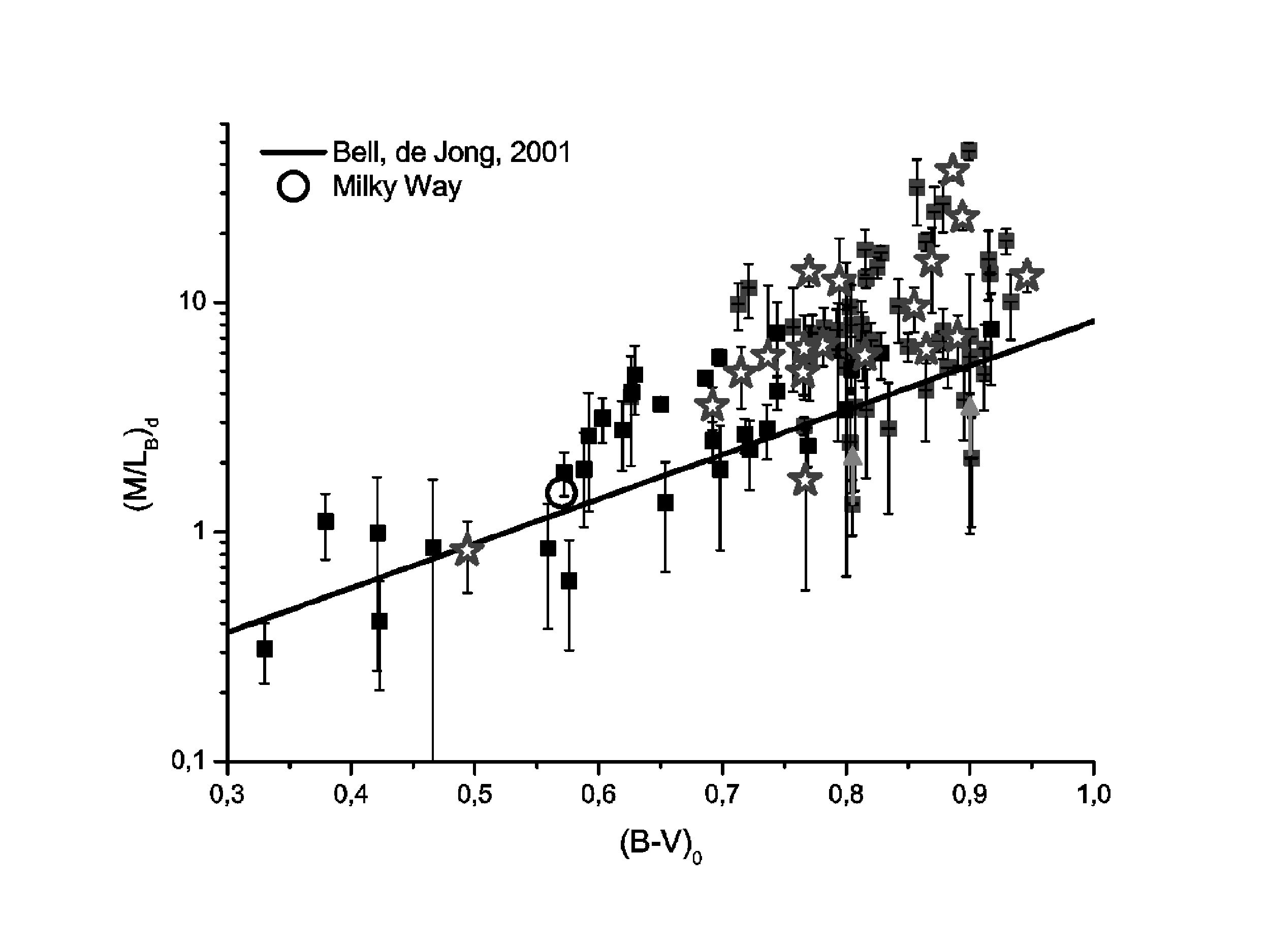}}
\caption{
\footnotesize
Mass-to-light ratio expected for stellar disks, which are marginally stable at $r\approx 2h$, over the color index of galaxies, in comparison with the ratios predicted by the evolutionary stellar model (straight line). Gray symbols correspond to the galaxies of S0-Sa types. Stars show the estimates based on the data obtained at the telescope BTA (SAO RAS).}
\label{fig1}
\end{figure}

\begin{figure}[]
\resizebox{7.5cm}{!}{\includegraphics{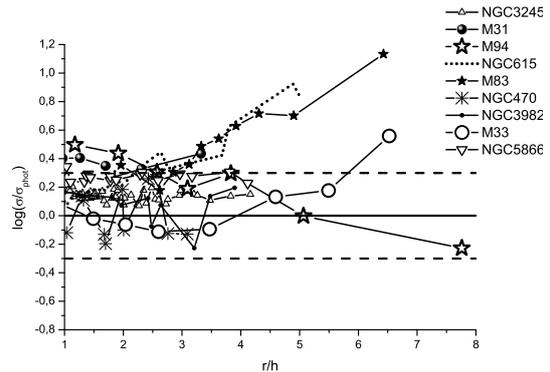}}
\caption{
\footnotesize
The ratio of the disk marginal surface density to the photometrically estimated one for different galaxies as a function of normalized radial distances $r/h$.}
\label{fig2}
\end{figure}

As it follows from \citet{Zasov2011} based on the available data analysis the disk surface density estimates at the radius $r=2h$ based on the marginal gravitational stability condition are close to that based on the photometry data together with the model ''M/L -- color'' relation for most of late type galaxies and for the significant fraction of of the lenticulars. It is demonstrated in Fig. \ref{fig1} where the mass-to-light ratios of stellar disks are plotted against the $(B-V)_0$ color index. Black line corresponds to the relations of \citet{bdj}. Gray symbols show the early type galaxies. The absence of systematic differences between the disk mass estimates based on the Toomre stability criterion and on the photometrical data for galaxies with $(B-V)_0<0.7$  indicates the absence of significant dynamical overheating for the majority of spiral galaxies and for some of lenticulars which lie along the straight line. It means that these galaxies have not experienced in their history a large number of merging events that could significantly dynamically heat the disk.
However the situation changes when we go to the radii larger than two disk radial scalelengths. In Fig. \ref{fig2} we demonstrate the ratio of the disk surface density $\sigma$, which corresponds to the marginally stable disk, to the density $\sigma_{phot}$ , resulting from photometric data (\citet{bdj} model) as a function of radial distance $r/h$ for galaxies with known velocity dispersion profiles. Photometric and kinematic data are taken from the literature. The most extended curves which belong to M33, M94 and M83 are obtained from planetary nebulae velocity dispersion measurements by \citet{HerrmannCiadullo} and \citet{Ciardullo}. The horizontal strip illustrates the range of the data uncertainty which is taken as a factor of two. From Fig. \ref{fig2} it is evident that in most cases the disks are slightly overheated or non-overheated, although in some cases $\sigma/\sigma_{phot}>2$ at large galactocentric distances. 

The problem of marginal gravitational stability of low surface brightness galaxies was also studied (\citet{Saburova2011}). According to this study the LSB disks may be marginally stable and in this case their surface densities appear to be significantly higher than it is expected from the photometry.

\section{Observations and data reduction}
\begin{figure}[]
\resizebox{6cm}{!}{\includegraphics{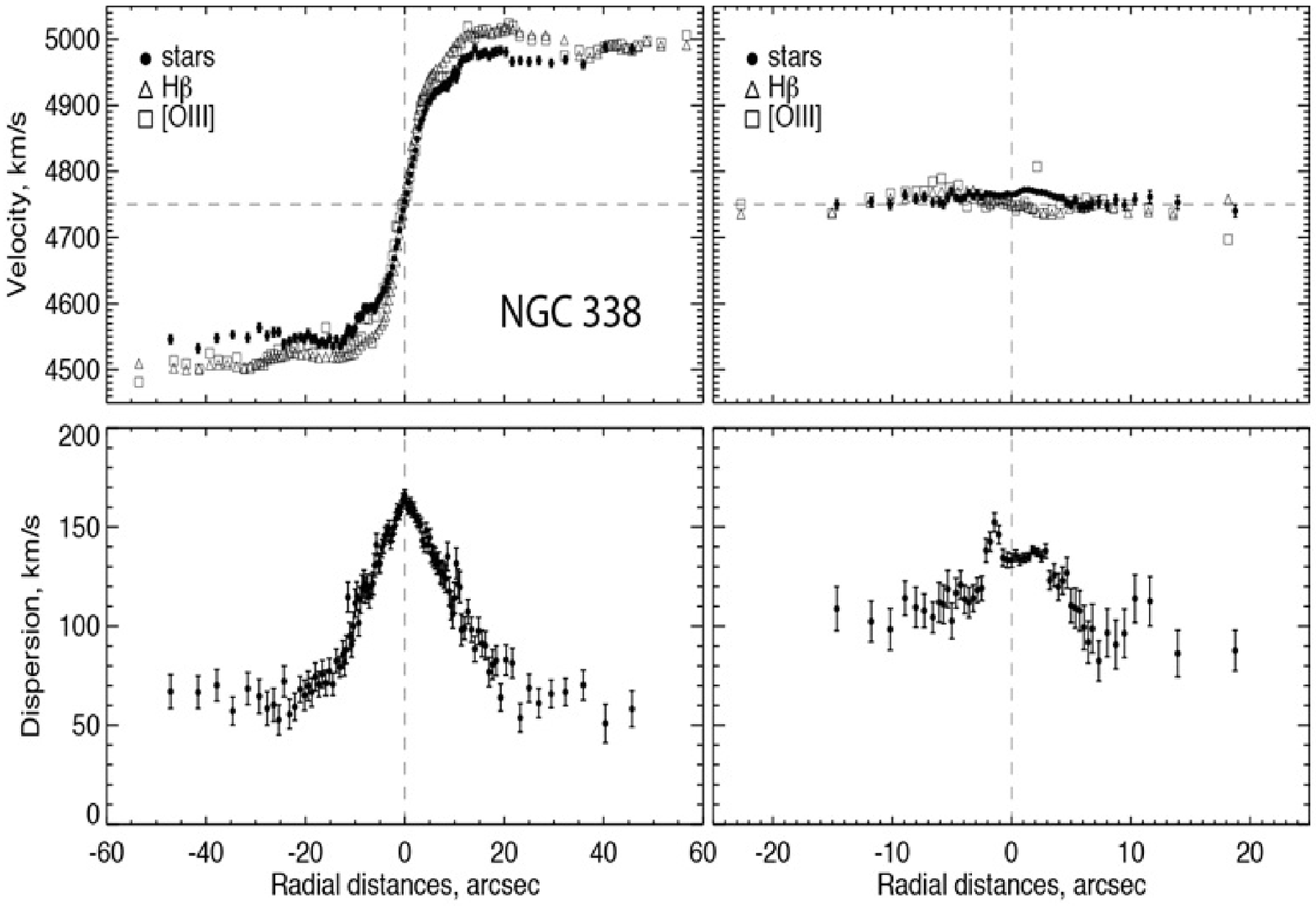}}
\resizebox{6cm}{!}{\includegraphics{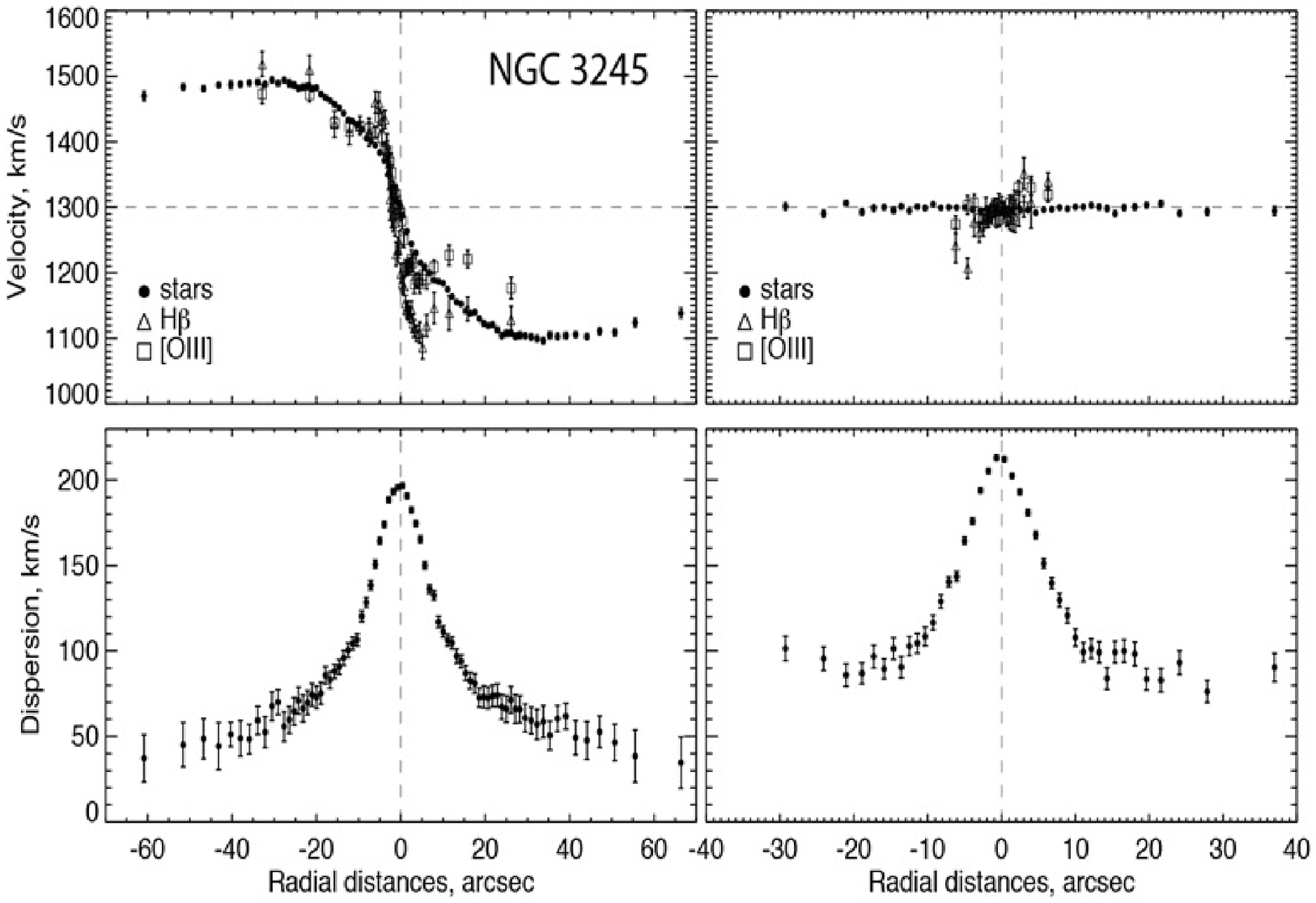}}

\caption{
\footnotesize
The examples of observed profiles of the line-of-sight velocity and velocity dispersion. The spectra were obtained along the major axis at BTA (SAO RAS) telescope.}
\label{fig3}
\end{figure}

Observations and data reduction were carried out in collaboration with the Special Astrophysical Observatory (SAO RAS), see \citet{Zasov2012}. The long-slit spectroscopic observations with the SCORPIO spectrograph (\citet{Afanasiev2005}) were used at the prime focus of the Russian 6-m BTA telescope.  The data reduction steps included the modeling of the spectral line spread function (LSF) variation  both along and across the wavelength direction (\citet{Katkov2011}) based on the  twilight spectrum. The parameters of internal kinematics and stellar populations of galaxies were obtained  by fitting high-resolution PEGASE.HR (\citet{Borgne2004}) simple stellar population  models using the NBURSTS full spectral fitting technique (\citet{Chilingarian2007a}A,B). 
The examples of the observed profiles of line-of-sight velocity and velocity dispersion of stellar disk along the major axis are given in Fig. \ref{fig3}. 
\section{Results of modeling.}

\begin{figure}[]
\resizebox{4cm}{!}{\includegraphics{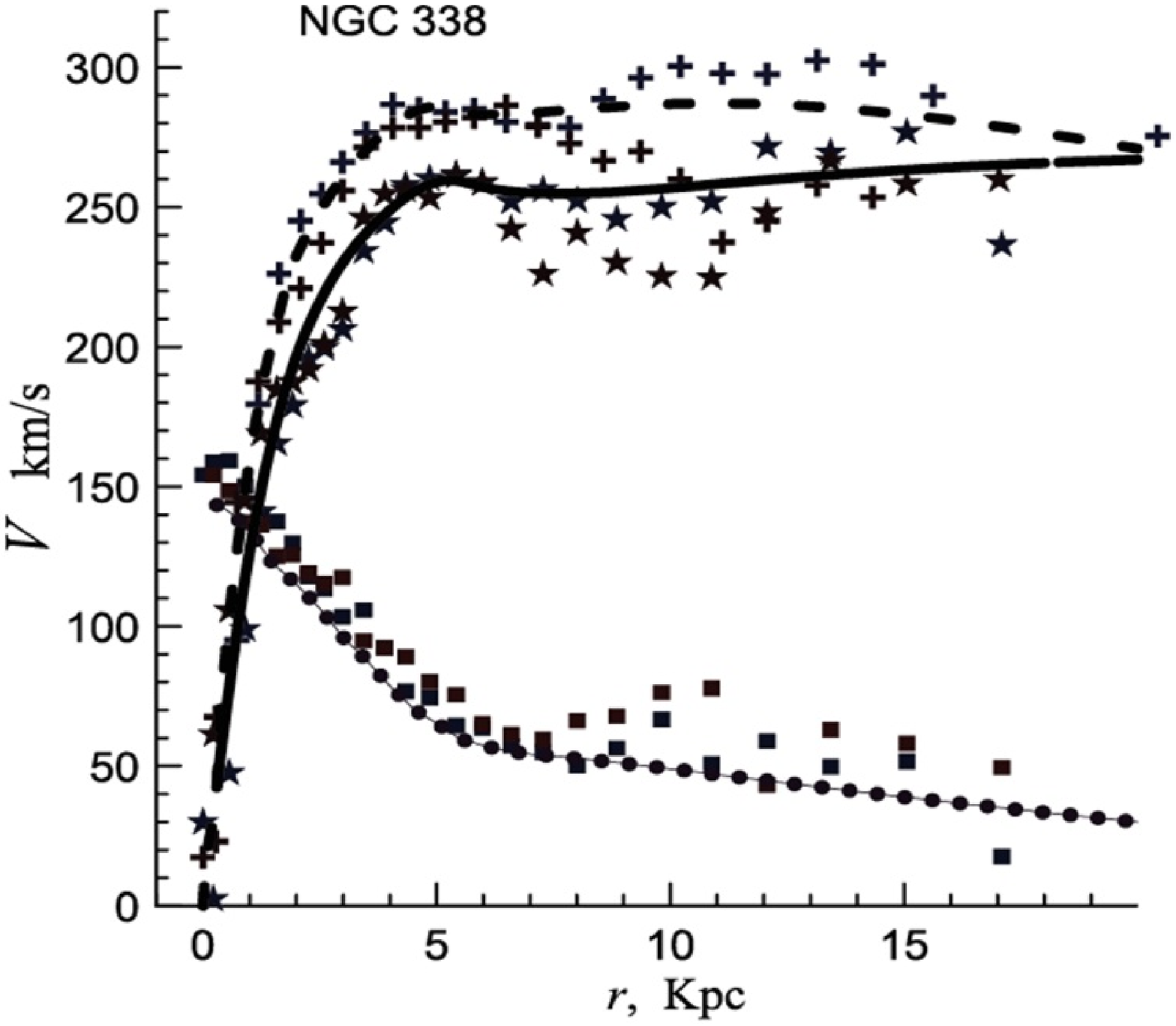}}
\resizebox{4cm}{!}{\includegraphics{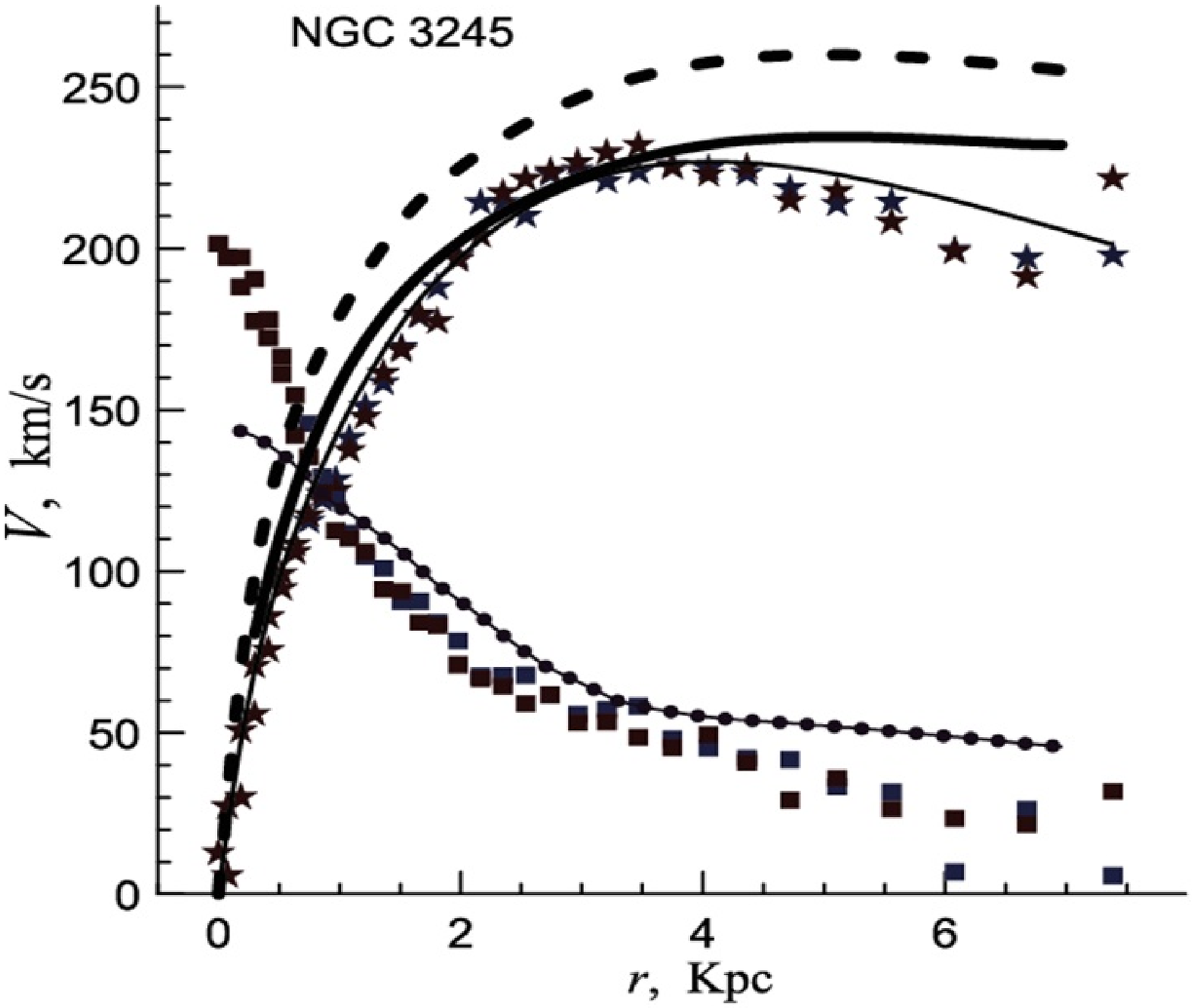}}

\caption{
\footnotesize
The model and observed radial profiles of velocity and velocity dispersion. Thick solid lines and dash lines  are the model rotation curves for stars and gas; asterisks and crosses are the observed de-projected velocities for stars and gas; squares and filled circles line (below) mark the radial runs of the observed stellar velocity dispersion and the maximal LOS dispersion expected for marginally stable disks respectively from \citet{Zasov2012}.}
\label{fig4}
\end{figure}

\begin{figure}[]
\resizebox{3.8cm}{!}{\includegraphics{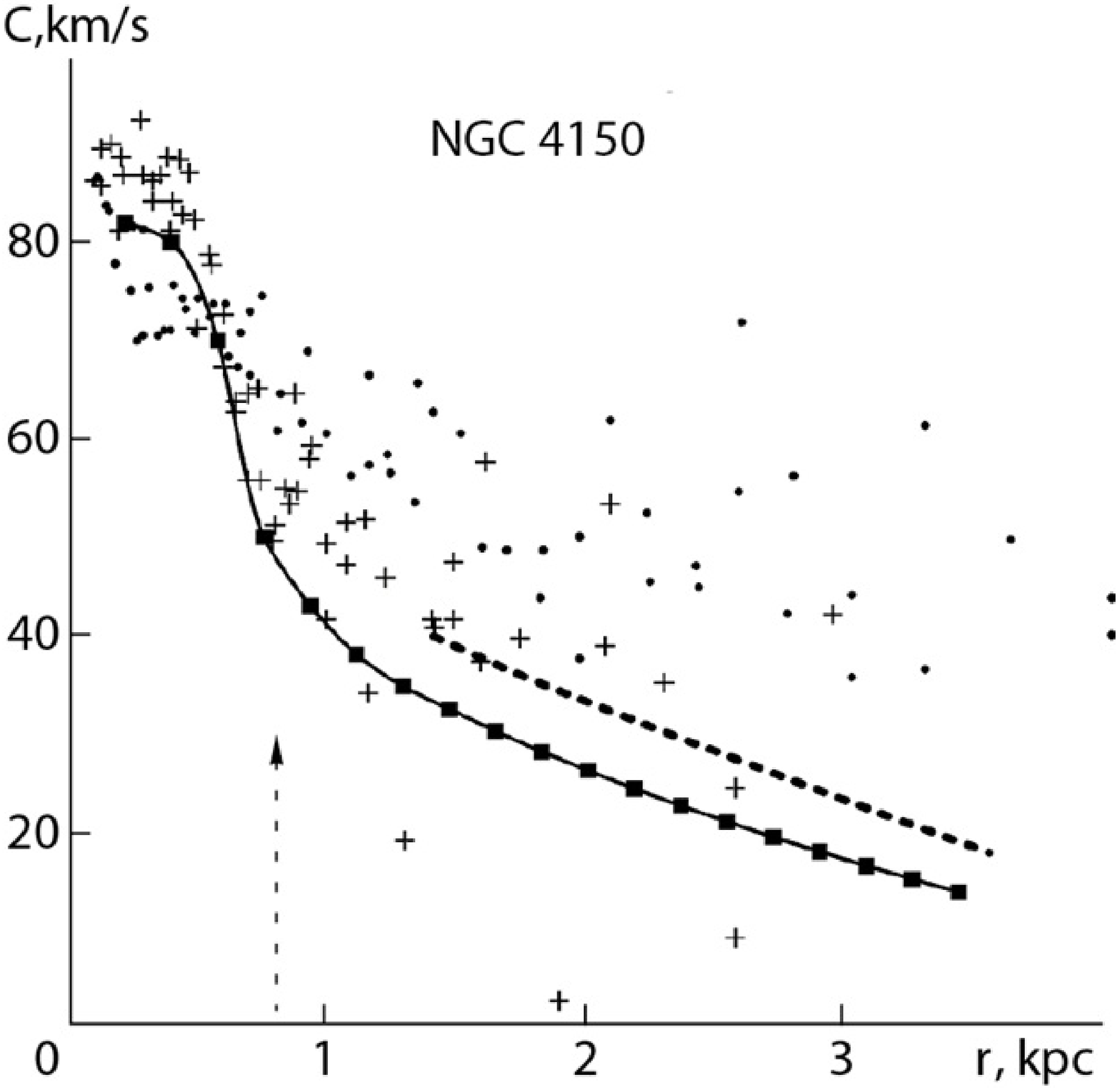}}
\resizebox{3.8cm}{!}{\includegraphics{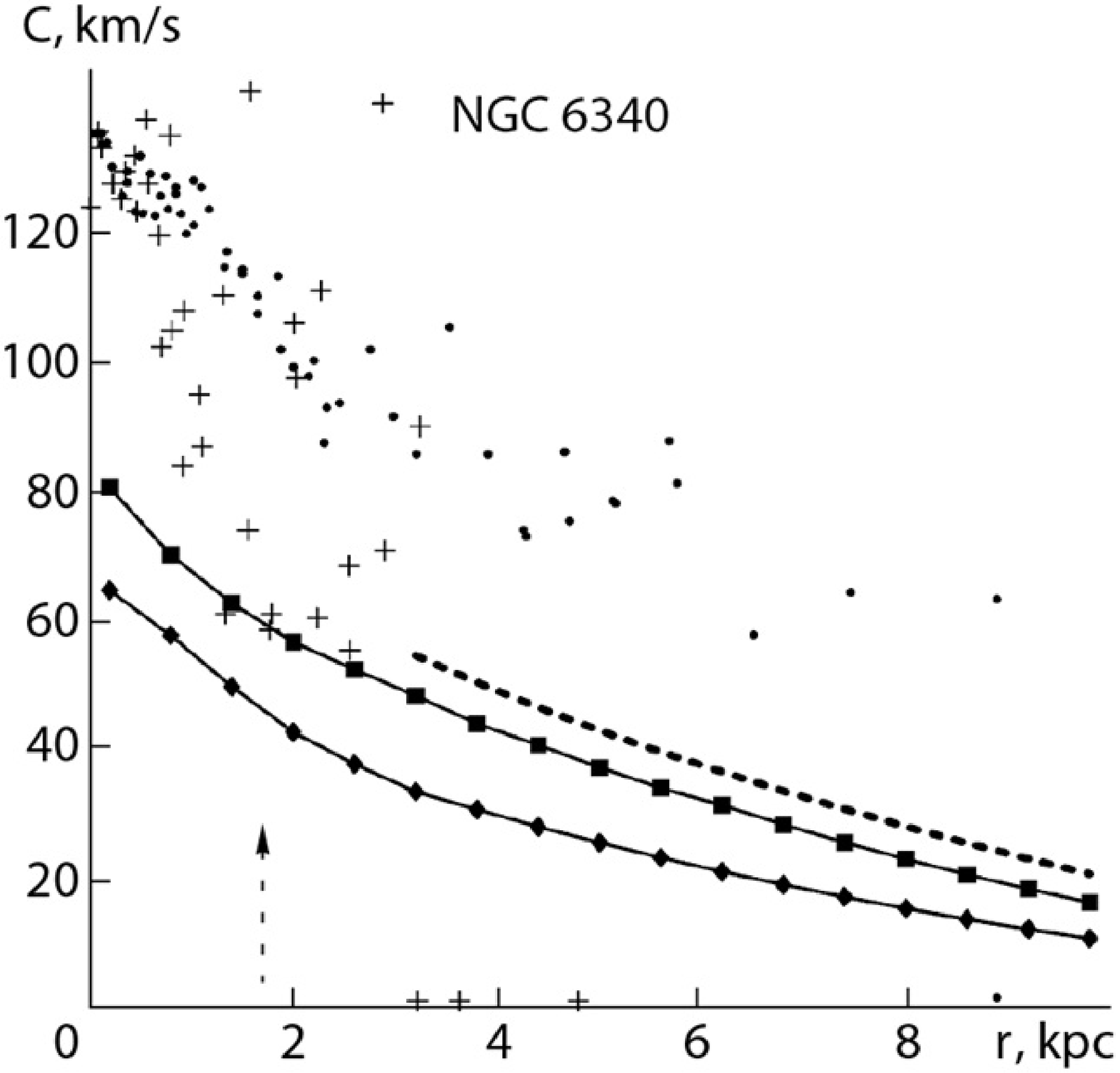}}

\caption{
\footnotesize
Stellar line-of-sight velocity dispersion observed along the major (crosses) and minor  (points) axes in comparison with the velocity dispersion expected for  the marginally stable disk - along the major (squares) and minor (dash curve) axes. Error bars ($\pm 10-15$ km/s) are not shown here \citet{Zasov2008}.}
\label{fig5}
\end{figure}

Fig. \ref{fig4} gives the example of the results of modeling of rotation curves and velocity dispersion profiles for two galaxies with marginally gravitationally stable disks \citet{Zasov2012}. 
However the situation is different for some of the galaxies which appear to be dynamically overheated. The model and observed velocity dispersion profiles are shown in Fig. \ref{fig5} for two galaxies with strongly overheated disks. 

Table \ref{t1} gives the results of the modeling for early type galaxies which were observed at BTA. As can be seen from the table, disks of some lenticulars are dynamically overheated although the others are consistent with marginal stability condition. It means that the lenticular galaxies may have different scenarios of the formation. Some of them experienced the merger events that significantly heated dynamically their disk while the others have managed to avoid such events.

\begin{table*}
\caption{Results for the early-type disky galaxies}
\label{t1}
\begin{center}
\begin{tabular}{lccccc}
\hline
\\
Galaxy & Type& Method & $M_{disk}$ obtained from the stability condition & Disk heating \\
\hline
\\
NGC338&S0&Modeling&$15.4\cdot 10^{10} M_{sun}$& Consistent with marginal stability\\
NGC524&S0& Analytical &$\ge 50\cdot 10^{10} M_{sun}$& overheated \\
NGC1167&S0& Modeling &$\ge 39\cdot 10^{10} M_{sun}$& overheated \\
NGC2273&SBa/S0&Modeling&$8.7\cdot 10^{10} M_{sun}$& Consistent with marginal stability\\
NGC3245&S0& Analytical &$5.1\cdot 10^{10} M_{sun}$& Consistent with marginal stability\\
NGC4124&S0& Analytical &$1.6\cdot 10^{10} M_{sun}$& Consistent with marginal stability\\
NGC4150&S0& Modeling &$\le 0.53\cdot 10^{10} M_{sun}$& overheated \\
NGC5440&S0& Modeling &$\le 19.3\cdot 10^{10} M_{sun}$& overheated \\
NGC6340&S0& Modeling &$\le 4.5\cdot 10^{10} M_{sun}$& overheated \\
\hline
\end{tabular}
\end{center}
\end{table*}
\section{Conclusions}
A dynamical heating of already formed stellar disks by external forces is not inevitable! 
Do mergers really play a role in the disk heating?\\
{\bf Spiral galaxies}: their thin stellar disks in most cases are close to marginally stable condition or slightly overheated within several radial scalelengths; \\
{\bf S0/Sa- galaxies}: there exist both galaxies with strongly overheated disks and galaxies with dynamically cool ones. The latters are mostly field galaxies which have avoided dynamical heating caused by interaction and merging. \\
{\bf Low surface brightness galaxies}: their extended disks may also be marginally stable (\citet{Saburova2011}). 

\begin{acknowledgements}
This work was partly supported by Russian Foundation for Basic Research, grants ofi11-02-12247, 12-02-00685, 12-02-31452
\end{acknowledgements}
\bibliographystyle{aa}

\begin{thebibliography}{}
\bibitem[Afanasiev \& Moiseev (2005)] {Afanasiev2005} Afanasiev, V.L.\& Moiseev, A.V., 2005, AstL, 31, 194

\bibitem[Bell \& de Jong (2001)] {bdj} Bell, E. F. \& 
de Jong, R.S., 2001, ApJ, 550, 212 

\bibitem[Chilingarian et al. (2007)A] {Chilingarian2007a} Chilingarian,  I.V., 2007A, MNRAS, 376, 1033 

\bibitem[ Chilingarian et al. (2007)B] {Chilingarian2007b} Chilingarian,  I. V., 2007B in Stellar Populations as Building Blocks of Galaxies, (IAU Symp. 241), p. 175

\bibitem[Ciardullo et al. (2004)] {Ciardullo} Ciardullo, R. et al., 2004, ApJ, 614, 167

\bibitem[Herrmann \& Ciardullo (2004)]{HerrmannCiadullo} Herrmann, K.A., Ciardullo, R., 2009, ApJ, 705, 1686

\bibitem[Katkov \& Chilingarian  (2011) ] {Katkov2011} Katkov, I.Y.,  Chilingarian, I.V., 2011, ASPC, 442, 143 

\bibitem[Khoperskov et al. (2003)] {Khoperskov2003} Khoperskov, A.V.,  et al. 2002, ARep, 47, 357 
\bibitem[ Le Borgne  et al. (2004)] {Borgne2004} Le Borgne, D. et al., 2004, A\&A, 425, 881 
\bibitem[Saburova (2011)]{Saburova2011} Saburova, A., 2011, ARep, 55, 409 

\bibitem[Zasov et al.(2011)] {Zasov2011} Zasov, A.~et al.\ 2011, AstL, 37, 374 

\bibitem[Zasov et al.  (2012) ] {Zasov2012} Zasov, A.V. et al., 2012, AstBu, 67, 353

\bibitem[Zasov et al. (2008)] {Zasov2008}
Zasov, A.~V. et al., 2008, ARep, 52, 79 







\end{thebibliography}

\end{document}